\documentclass{pasa}

\def\Curtin{$^{1}$}
\def\CAASTRO{$^{2}$}
\def\INAF{$^{4}$}
\def\ASU{$^{5}$}
\def\Victoria{$^{7}$}
\def\Peripety{$^{8}$}
\def\UWisc{$^{9}$}
\def\UW{$^{10}$}
\def\UWA{$^{11}$}

\newcommand{\newtext}[1]{{#1}}

\title[The Engineering Development Array]{The Engineering Development Array: A low frequency radio telescope utilising SKA precursor technology}
\author[Wayth et al.]{Randall~Wayth\Curtin$^,$\CAASTRO\thanks{r.wayth@curtin.edu.au},
Marcin Sokolowski\Curtin$^,$\CAASTRO,
Tom Booler$^{1}$,
Brian Crosse$^1$,
David Emrich$^1$,
Robert Grootjans$^{3,1}$,
Peter J. Hall$^{1}$,
Luke Horsley$^1$,
Budi Juswardy$^{1}$,
David Kenney$^1$,
Kim Steele$^1$,
Adrian Sutinjo$^{1}$,
Steven~J.~Tingay\Curtin$^,$\CAASTRO$^,$\INAF, 
Daniel Ung$^{1}$,
Mia Walker$^1$,
Andrew Williams$^{1}$,
A.~Beardsley\ASU,
T.~M.~O.~Franzen\Curtin,
M.~Johnston-Hollitt\Victoria$^,$\Peripety,
D.~L.~Kaplan\UWisc, 
M.~F.~Morales\UW, 
D.~Pallot\UWA,
C.~M.~Trott\Curtin$^,$\CAASTRO,
C.~Wu\UWA
\\
\affil{$^1$International Centre for Radio Astronomy Research (ICRAR), Curtin University. GPO Box U1987, Perth, 6845. Australia}%
\affil{$^{2}$ARC Centre of Excellence for All-sky Astrophysics (CAASTRO)}%
\affil{$^3$University of Twente, PO Box 217, 7500 AE Enschede, Netherlands.}%
\affil{$^{4}$Istituto Nazionale di Astrofisica (INAF) -- Istituto di Radio Astronomia, Via Piero Gobetti, Bologna, 40129, Italy}%
\affil{$^{5}$School of Earth and Space Exploration, Arizona State University, Tempe, AZ 85287, USA}%
\affil{$^{6}$International Centre for Radio Astronomy Research, Curtin University, Bentley, WA 6102, Australia}%
\affil{$^{7}$School of Chemical \& Physical Sciences, Victoria University of Wellington, P.O. Box 600 Wellington 6140, New Zealand}%
\affil{$^{8}$Peripety Scientific Ltd., P.O. Box 11355 Manners Street, Wellington 6140, New Zealand}%
\affil{$^{9}$Department of Physics, University of Wisconsin--Milwaukee, Milwaukee, WI 53201, USA}%
\affil{$^{10}$Department of Physics, University of Washington, Seattle, WA 98195, USA}%
\affil{$^{11}$International Centre for Radio Astronomy Research (ICRAR), University of Western Australia, Crawley 6009, Australia}%
}%
\jid{PASA}
\doi{10.1017/pas.\the\year.xxx}
\jyear{\the\year}

\usepackage[authoryear]{natbib}
\bibpunct{(}{)}{;}{a}{}{,}
\usepackage{aas_macros}
\usepackage{hyperref} 
\hypersetup{colorlinks,citecolor=blue,linkcolor=blue,urlcolor=blue}
\usepackage{gensymb} % MS added
\usepackage{caption} % MS added to compile on older version - can be removed if does not work on new version now ...
\usepackage{subcaption}

\begin{document}%
\begin{abstract}
We describe the design and performance of the Engineering Development Array (EDA), which is a low frequency radio telescope comprising 256 dual-polarisation dipole antennas working as a phased-array.
The EDA was conceived of, developed, and deployed in just 18 months via re-use of Square Kilometre Array (SKA) precursor technology and expertise, specifically from the Murchison Widefield Array (MWA) radio telescope.
Using drift scans and a model for the sky brightness temperature at low frequencies, we have derived the EDA's receiver temperature as a function of frequency. The EDA is shown to be sky-noise limited over most of the frequency range measured between 60 and 240\,MHz.
By using the EDA in interferometric mode with the MWA, we used calibrated visibilities to measure the absolute sensitivity of the array. The measured array sensitivity matches very well with a model based on the array layout and measured receiver temperature.
The results demonstrate the practicality and feasibility of using MWA-style precursor technology for SKA-scale stations.
The modular architecture of the EDA allows upgrades to the array to be rolled out in a staged approach.
Future improvements to the EDA include replacing the second stage beamformer with a fully digital system, and to transition to using RF-over-fibre for the signal output from first stage beamformers.

\end{abstract}
\begin{keywords}
telescopes -- instrumentation: miscellaneous
\end{keywords}
\maketitle%
\section{INTRODUCTION}
\label{sec:intro}
%- Background: SKA science\\
The recent renaissance in low frequency radio astronomy is being driven by high priority goals in modern cosmology \citep[e.g. ][]{2015aska.confE...1K}. The new and upgraded radio telescopes pursuing this science are predominantly dipole-based arrays.
In the case of the Murchison Widefield Array \citep[MWA, ][]{2013PASA...30....7T} and the Low Frequency Array \citep[LOFAR, ][]{2013A&A...556A...2V}, the antennas form small aperture array `tiles' as the fundamental receptor elements in the system. 
Other arrays such as
the Precision Array for Probing the Epoch of Reionization \citep[PAPER, ][]{2010AJ....139.1468P},
the Long Wavelength Array \citep[LWA, ][]{2013ITAP...61.2540E},
and the Large-Aperture Experiment to Detect the Dark Ages \citep[LEDA, ][]{2012arXiv1201.1700G} have the dipoles themselves as the fundamental receptor elements, hence are true all-sky instruments.
Exceptions to all-dipole array exist, of course, notably the upgraded Giant Metrewave Radio Telescope \citep[GMRT, ][]{2014ASInC..13..441G} and
Karl G. Jansky Very Large Array \citep[JVLA, ][]{2011ApJ...739L...1P} telescopes,
and the under-construction Hydrogen Epoch of Reionization Array \citep[HERA, ][]{2017PASP..129d5001D} telescope.

%- relationship to SKA\\
The current design of the Square Kilometre Array Low Frequency array  \citep[SKA-Low, ][]{Turner2015} calls for low frequency antenna elements arranged in a pseudo-random layout, working as a phased array `station', as the fundamental receptor element for most observing modes.
SKA-Low stations are 35\,m in diameter consisting of 256 antennas.
The chosen size and sensitivity of the SKA-Low stations is the result of a complex trade-off between system complexity and many competing science requirements. 

%- purpose: test and verification platform \\
The Engineering Development Array (EDA) was conceived as a rapidly deployable test and verification platform to support SKA Low work.
The high-level goals of creating the EDA were:
\begin{itemize}
\item to create a reference platform to the proposed SKA Low stations, using existing well understood components;
\item to explore a risk mitigation path for SKA Low, based on well understood and robust antenna and beamforming solutions;
\item to enable the enhancement of the MWA correlator \citep{2015PASA...32....6O} to accept data from external instruments, in anticipation of integration with the future SKA Low prototype Aperture Array Verification System 1 (AAVS1);
\item to enable prototyping and development work for potential future MWA upgrades, and
\item to provide an easily accessed experimental platform for validating new-generation sparse array design and metrology techniques, with application to the SKA and other radio science applications.
\end{itemize}
Our experience with the ``AAVS0.5'' test system \citep{2015ITAP...63.5433S}, which was fully integrated with the MWA, showed that the re-use of an existing interferometer (including software and down-stream processing tools) enabled the rapid, accurate characterisation of the prototype SKALA antennas \citep{2015ExA....39..567D}.
It was also recognised from the outset that a facility like the EDA could be used for specialised science investigations in a standalone mode, as well as research into techniques using hybrid antenna systems for low frequency radio interferometers when combined with the MWA or AAVS1.

The EDA is sited at the Murchison Radio-astronomy Observatory (MRO) in Western Australia and is hosted by the MWA under the MWA's external instrument policy\footnote{see \url{http://www.mwatelescope.org/team}}.
The EDA has similarities to LOFAR stations and LWA stations, but is functionally and conceptually more closely related to the proposed SKA Low station, by design.
%- Similar instruments\\

\section{DESCRIPTION}
%- detailed design\\
\label{sec:desc}
\begin{figure}
\begin{center}
\includegraphics[width=\columnwidth]{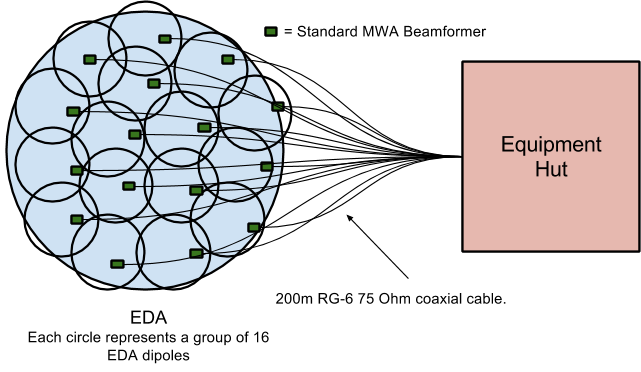}
\caption{The high-level design of the EDA's signal path}
\label{fig:high_level_design}
\end{center}
\end{figure}

The EDA consists of antenna hardware, beamformers, data capture and signal processing components. The essential properties of the EDA are listed in Table \ref{tab:specs}.
The antenna system consists of 256 dual-polarisation dipoles pseudo-randomly distributed over an area 35\,m in diameter.
An overarching design principle for the EDA was to re-use as much existing hardware and software as possible, to reduce development time and risk.

The EDA's antenna elements are standard MWA dipoles that have slightly modified low-noise amplifiers (LNAs) to extend the frequency range of the received signals down towards 50\,MHz. The dipoles are clipped to a groundscreen mesh that was aligned with local true north-south and east-west cardinal directions by professional surveyors.
\newtext{The actual layout of the dipoles was chosen to be identical to the proposed AAVS1 system, where the locations of the dipoles were chosen randomly with the only constraints that they be inside the 35\,m station, and no closer than 1.5\,m to another dipole.}

\begin{table}
\caption{Specifications of the EDA}
\begin{center}
\begin{tabular}{@{}lr@{}}
\hline%
 Number of dipoles  & 256 \\
 Number of in-field beamformers & 16 \\
 Dipoles per beamformer  & 16 \\
 Diameter & 35\,m \\
 Location(lon,lat degs) & 116.672257,-26.703051 \\
 Frequency range (MHz) & 50 to 300 \\
\hline\hline
\end{tabular}
\end{center}
\label{tab:specs}
\end{table}

As with the MWA, the dipoles are connected in groups of 16 to standard MWA analogue beamformers, hence there are 16 beamformers servicing all 256 dipoles (Fig. \ref{fig:high_level_design}).
The beamformers provide power to the dipoles and combine the output signals from the dipoles with a switchable true time delay for each dipole that is an integer multiple of 435\,ps, where the multiplier ranges from 0 to 31.
%In this way the beamformers can point each sub-array of 16 dipoles on the sky.
Each beamformer produces two outputs, which are the combined signals from its 16 dipoles, one for each polarisation.
Details of the MWA dipoles and beamformers can be found in \citet{2009IEEEP..97.1497L,2013PASA...30....7T}.

\begin{figure}
\begin{center}
\includegraphics[width=\columnwidth]{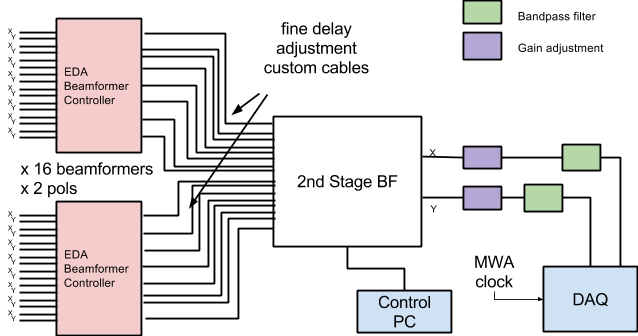}
\caption{The signal path inside the equipment hut. The two EDA beamformer controller units each house eight standard MWA ``Data over coax'' (DoC) cards, which send power to fielded beamformers, provide digital communications, and pass RF signal on to downstream components.}
\label{fig:inside_hut}
\end{center}
\end{figure}

The 16 beamformers are connected via 200\,m of low-loss RG-6 quad-shield coaxial cable to beamformer-controller units housed in an RF-shielded equipment room (the ``Hut'') near the EDA and MWA's infrastructure hub.
The beamformer-controller units provide both power to, and serial communications with, the beamformers via the coaxial cables and pass the RF signals to the second stage beamformer.
Each beamformer-controller unit contains a power supply (supplying power to the eight connected first stage beamformers), a custom-built mainboard, and eight standard MWA ``Data over Coax'' (DoC) interface cards all housed in a standard 4RU enclosure.
The mainboard hosts two Raspberry Pi single-board computers as well as serial communication interface logic, power control and monitor circuitry.
One single-board computer controls and monitors power and temperatures, while the other receives pointing commands and communicates these to the beamformers via the DoC cards.
The enclosure and mainboard for the beamformer-controllers represent the only significant in-house hardware development that was required for the EDA.
The output of the beamformer-controllers are the RF signals from each first stage beamformer, which are passed to the second stage beamformer.
Second stage beamforming is performed in the Hut (see Fig \ref{fig:inside_hut}).

%- figure showing dipole layout, beamformer associations.\\
\begin{figure}
\begin{center}
\includegraphics[width=\columnwidth]{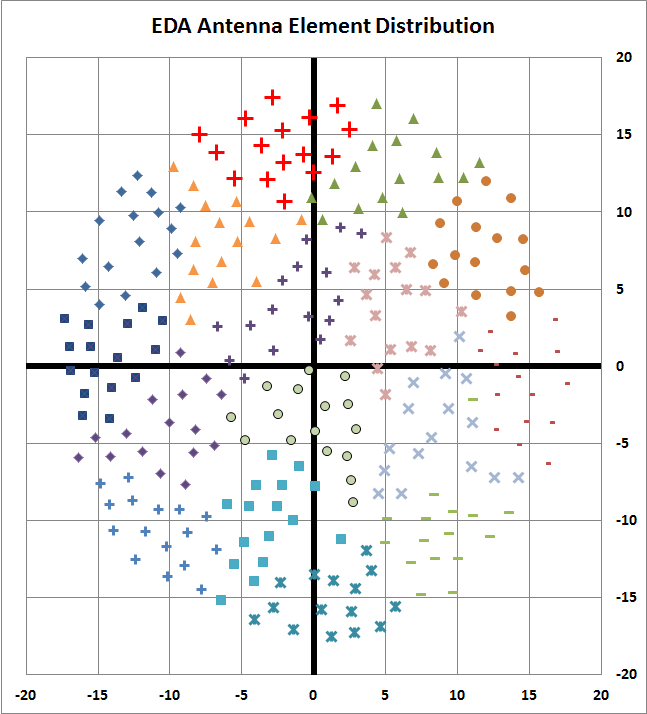}
\caption{The pseudo-random layout of the EDA's dipoles. North is up, east to the right, distances in meters. The dipole symbols denote how they are grouped for the analogue beamformers.}
\label{fig:dipole_layout}
\end{center}
\end{figure}

The MWA beamformers were designed for MWA tiles, which are physically more compact than the sub-arrays used in the EDA or in AAVS0.5.
The fraction of sky accessible by the EDA is correspondingly limited by the maximum delay of an MWA beamformer, which is 13.5\,ns.
Fig. \ref{fig:dipole_layout} shows that the typical size scale of one beamformer's sub-array in the EDA is 10\,m, which limits the maximum zenith angle for any one sub-array to around 25 degrees.
Zenith angles greater than 25 degrees can be reached by simply disabling dipoles whose required delays exceed the maximum achievable, with the sensitivity of the array degrading with increased zenith angle.
To achieve pointing zenith angles up to approximately 60 degrees with the full array, the maximum delay in the first stage analogue beamformers must be increased to around 28\,ns.

\newtext{We note that since the MWA beamformers are true time delay beamformers with independent delay settings for each input, there is no restriction on how dipoles can be arranged within a sub-array.}

\subsection{Second-stage beamforming}
In the second stage beamformer, signals from all external MWA beamformers are combined, with appropriate delays, to form a phased-array signal from the entire array.
The outputs of the second-stage beamformer are also two signals, one for each polarisation of the dipoles.
The initial implementation of the second stage beamformer uses a separate prototype analogue beamformer (the ``Kaelus'' beamformer), which was commissioned on behalf of the SKA Low Aperture Array Design and Construction (AADC) consortium as a potential alternative design to full digital beamforming for SKA Low stations. This option was eventually not adopted by the AADC, however the prototype unit was suitable as a fast and low-risk way to complete the EDA.
In parallel with this, development work for digital second stage beamforming was undertaken.

\begin{figure*}
\begin{center}
\includegraphics[width=\textwidth]{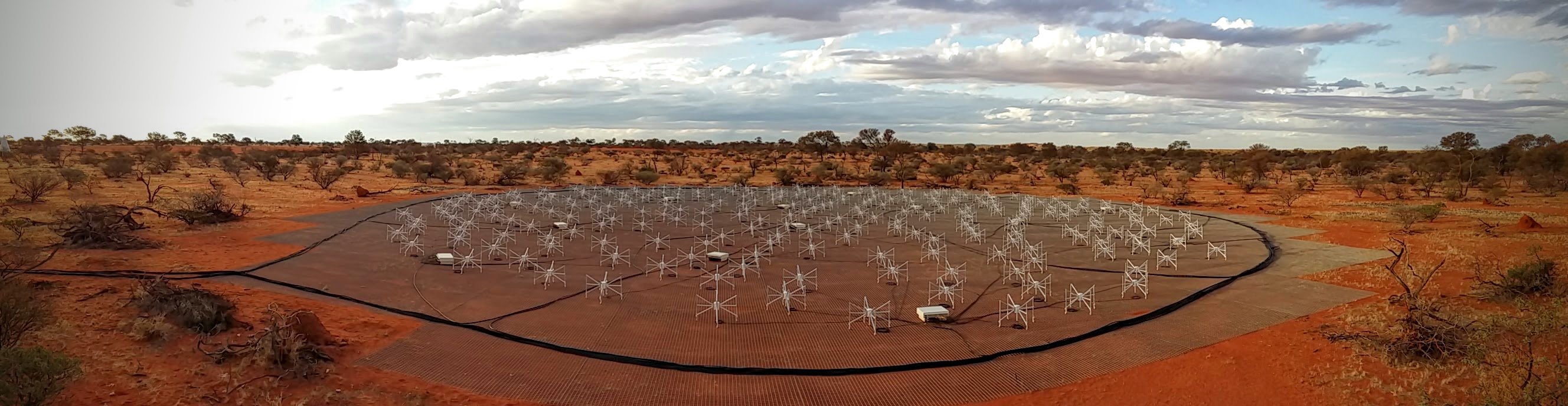}
\caption{A panorama of the EDA looking north. In-field beamformers (small white boxes) each service 16 dipole antennas. The thick black cables are the 200\,m coaxial cables running between the beamformers and the shielded room.}
\label{fig:panorama}
\end{center}
\end{figure*}

Prior to deployment, careful attention was paid to measuring and correcting for electrical length differences between the 16 pairs of long coaxial cable that run between the first stage beamformers and the Hut. The small differences in lengths of the 200\,m cables were corrected using custom length connecting cables inside the Hut (see Fig \ref{fig:inside_hut}) such that the total electrical length of cable between the output of the first stage beamformers and the input of the second stage beamformer was equal to better than 2\,cm.
A photo of the deployed EDA is shown in Figure \ref{fig:panorama}.

The use of an analogue beamformer for the second stage beamforming introduces some reduction in performance as discussed below.
In addition, the maximum delay available in the Kaelus second stage beamformer limits the maximum zenith angle to around 20 degrees (it was originally designed as a first stage beamformer), however that level of sky coverage is adequate for characterisation of the system performance.
The sky coverage limitation and performance degradation issues will be removed with the use of digital second stage beamforming.
Examples of the array beam formed by the EDA at zenith are shown in Figure \ref{fig:EDA_beams}.

\newtext{The phase centre of the EDA is defined to be the centre of the station at coordinate (0,0) in Figure \ref{fig:dipole_layout}.}
To point the array, the required geometric delays are calculated for each dipole in the array and delays are apportioned between first and second stage beamformers such that the sum of the squares of delay error over all dipoles is minimised. 
\newtext{In this scheme, it is not necessary to define a phase centre for each sub-array, however in practise the sub-array phase centre is constrained to be close to the physical centre. Since the beamformers can only implement zero or positive delays, we define the ``zero delay'' state of both beamformers to be half the maximum available delay and use values less than half the maximum for negative delays relative to the phase centre and values greater than half the maximum for positive.}
%- limitations of analogue beamforming\\

\subsection{Data capture and digital systems}
\label{sec:datacapture}
The output of the second stage beamformer is lowpass filtered and sampled by a Signatec PX1500-2 commercial data acquisition card, housed in a Supermicro X9DR3-LN4+ server.
The card's sample clock is directly driven by an MWA clock distribution unit, hence it samples the RF signal at 655.36\,Msamp/sec coherently with the MWA.
8-bit samples from both polarisations are transferred to an nvidia GeForce GTX 750 graphics processing unit (GPU) in the same server, where they are directly transformed to 10\,kHz channels (to match the MWA) via a 65536-sample FFT using the CUDA cuFFT library.
This single stage filterbank thus generates the same fine channel resolution as the two-stage filterbank used in the MWA \citep{2015ExA....39...73P,2015PASA...32....6O}  without any gaps or artefacts in the spectrum.

% see values in eda_beamformer_swap_test_AFTER_AW_CHANGE_FAST_TEST_20161123.odt and ./picosec2meters.py 500 or 7000
% phases -> electrical lengths
After deployment, the total electrical length of each sub-array (from antennas to digitiser input) was measured by turning off 15 of the 16 sub-arrays in the EDA and measuring the delay relative to an MWA reference antenna using an astronomical source (3C444 and Hydra A).
This test showed that the as-deployed sub-arrays had end-to-end electrical length differences up to 20\,cm.
%{The reason for the extra length difference is not clear, but could be related to small changes introduced when unrolling the cables from their spools.}
Therefore, the extra delays (different for each of the 16 beamformers) were tabulated and added to delays calculated for every pointing.
This adjustment reduced the standard deviation of the phase error between the 16 beamformers at 200-230\,MHz to below 5 degrees.
% These extra delays were measured with respect to average phase. 

\subsection{Integration with the MWA}
The data sampled from the EDA via the PX1500-2 card is fully coherent with the MWA since it uses the shared MWA clock.
To enable correlation of EDA data with the MWA, a mechanism was created to select 3072 of the 10\,kHz fine channels (corresponding to 24 sets of 128 fine channels generated by the MWA's digital systems) from the EDA's filterbank output.
The original plan for integration with the MWA was to ingest the EDA data into the correlator and replace a datastream from an existing antenna.
This method uncovered an unexpected problem when combining digital data that have been generated with heterogenous filterbank systems, a detailed description of which will appear in another paper.

Instead, to enable the sensitivity measurements described below, the output of the second stage beamformer was temporarily connected directly to the nearest MWA receiver, physically replacing the input connections from an MWA tile. In this way the EDA's signal looked like just another MWA tile \citep[as was done in ][]{2015ITAP...63.5433S} and the data capture and processing proceeded via normal MWA tools.

\begin{figure*}
\hspace{-10mm}
\centering
    \begin{subfigure}[b]{0.33\textwidth}
                \label{fig:pb_69}
                \includegraphics[width=1.0\textwidth]{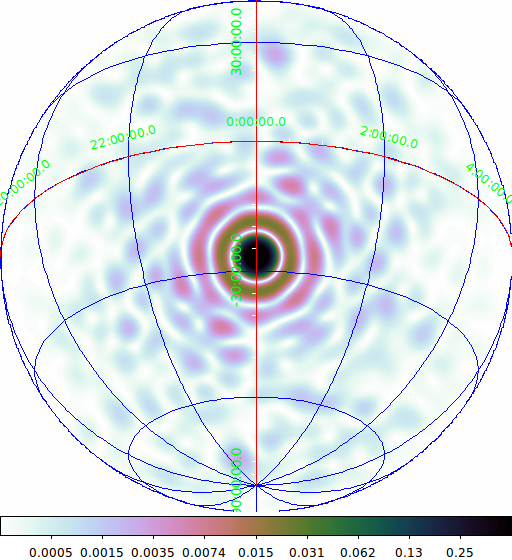}
		 \caption{100 MHz}
    \end{subfigure}
    \begin{subfigure}[b]{0.33\textwidth}
                \label{fig:pb_93}
                \includegraphics[width=1.0\textwidth]{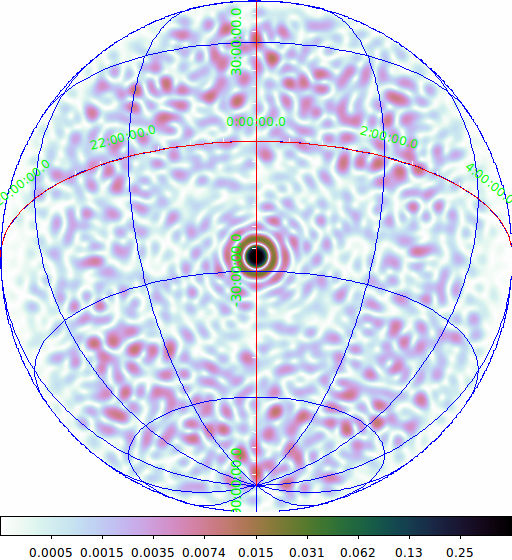}
		 \caption{200 MHz}
    \end{subfigure}
    \begin{subfigure}[b]{0.33\textwidth}
                \label{fig:pb_121}
                \includegraphics[width=1.0\textwidth]{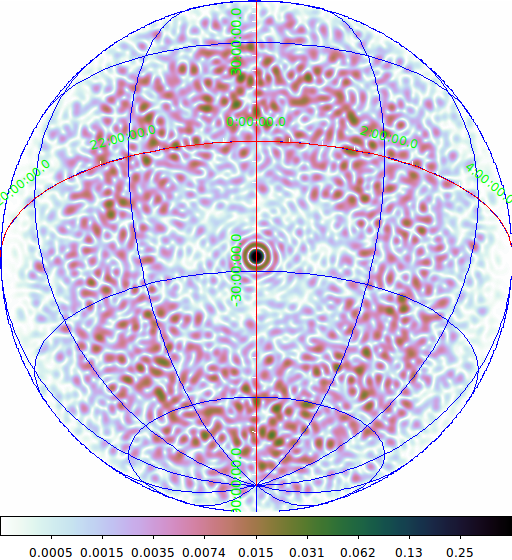}
		 \caption{300 MHz}
    \end{subfigure}

    \caption{Examples of simulated EDA array beams (in linear power relative to the maximum) pointed at the zenith for the east-west oriented dipoles.}
    \label{fig:EDA_beams}
\end{figure*}

%%%%%%%%%%%%%%%%%%%%%%%
%%%%%%%%%%%%%%%%%%%%%%%
\section{PERFORMANCE}
\label{sec:performance}
The theoretical performance of the EDA is determined entirely by the performance of the individual dipole elements and the station layout. 
In practise, natural component tolerances will generate amplitude and phase differences that will reduce sensitivity compared to ideal.
To first order, if each dipole has amplitude gain $1+\sigma_{a}$ and phase error $\sigma_{p}$ radian, where $\sigma$ is the standard deviation calculated over all dipoles, then the reduction in the station beam directivity will be
\begin{equation}\label{eq:amp_sigma}
D/D_{\mathrm{ideal}} = \frac{1}{1+\sigma_{a}^{2} + \sigma_{p}^{2}},
\end{equation}
where $D_{\mathrm{ideal}}$ is the ideal case where all gains are equal.
%\emph{Adrian: please check this bit is correct.}
In general, amplitude errors are frequency independent whereas phase errors in this system are caused by delay mismatches, hence are frequency dependent.

The gain variation of MWA LNAs are meant to be small by design.
\citet{2016ApJ...820...44N} measured a $1\sigma$ variation of 0.09\,dB around the nominal gain of 19\,dB (i.e. $2.3\%$ variation around a linear gain of 79.4) and negligible gain variation in the coaxial cables that join the dipoles to the beamformers (again, by design).
Prior to deployment, the standard deviation in the gain amplitude of the 16 MWA beamformers was measured to be 0.7\,dB.
Since the outputs of beamformers in the regular MWA are digitised, and hence gain variations are absorbed into the normal calibration of the array, it was never a requirement for the gain amplitudes between beamformers to be tightly specified.
Hence the larger amplitude variations seen in beamformer gain is not surprising.

Since the EDA as initially deployed uses analogue second stage beamforming, the amplitude variations between first stage beamformer gains will cause reduction in the array directivity, which is taken into account in the following analysis.

\subsection{Expected Performance}
\label{sec:expectedperf}
%- table of typical RMS values (gain and/or phase) of analogue components for spot frequencies\\
%- expected degradation in beamforming performance\\
%- discussion of deviation of analogue components from ideal performance\\
According to Equation (\ref{eq:amp_sigma}), gain variations of $2.3\%$ in the LNAs should cause less than $0.1\%$ loss in theoretical sensitivity of the array.
We therefore expect that it is only the uncorrected amplitude gain variations between first stage beamformers, and the residual phase variations that cause loss of sensitivity compared to the ideal.
The 5 degree (0.09 radian) residual phase errors in the 200-230\,MHz range produce a $\sim 1\%$ reduction in theoretical sensitivity.
The 0.7\,dB gain variation measured between the first stage beamformers (which have a nominal gain around 40\,dB) corresponds to a 15\% $1\sigma$ variation in linear units. Hence, we expect this uncorrected gain amplitude variation to cause at least 3\% loss in theoretical sensitivity.
We note also that since all dipoles connected to a first stage beamformer experience the same gain/delay offset, the errors caused by the analogue second stage beamformer are partially correlated between different dipoles. Hence these loss estimates should be considered a lower bound.

\begin{figure}
\begin{center}
\includegraphics[width=0.5\textwidth]{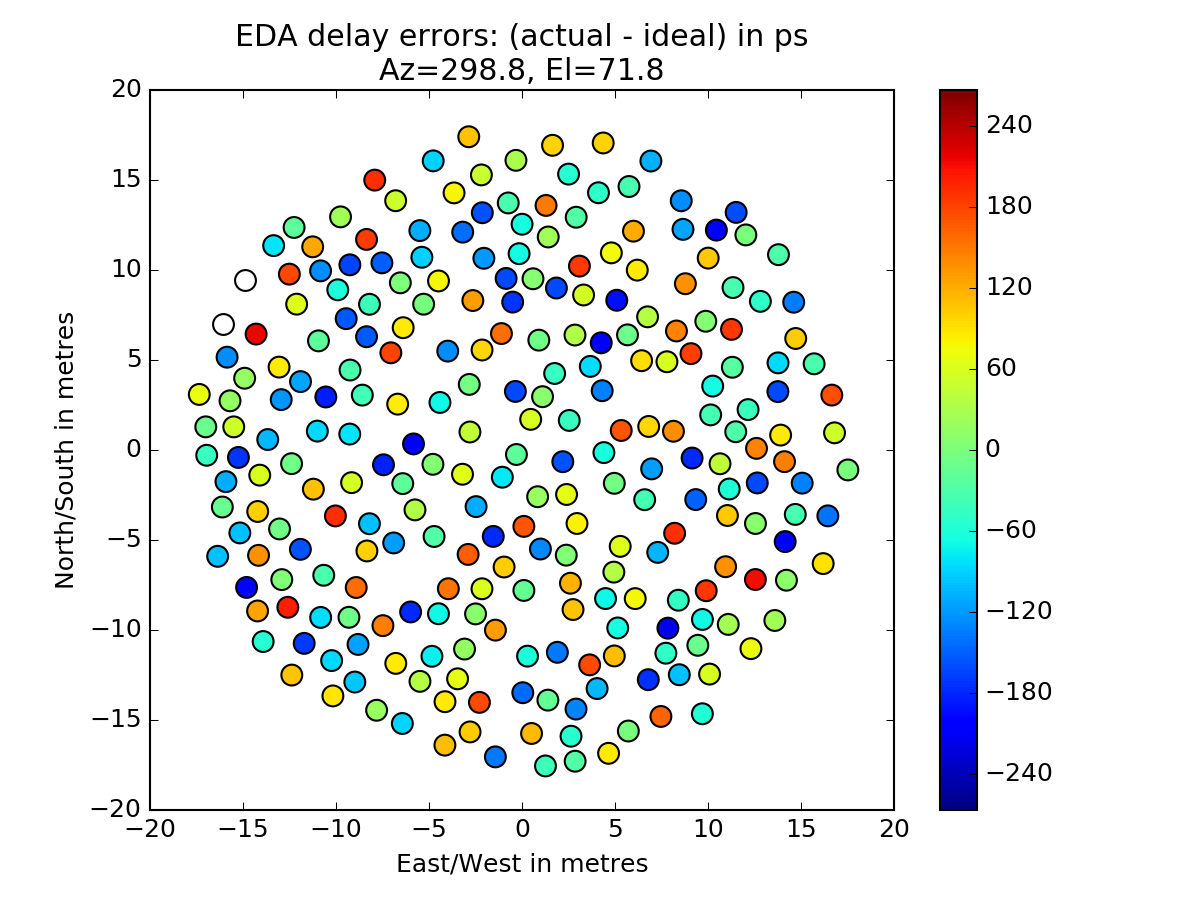}
\caption{An example of the theoretical instantaneous delay errors introduced on each dipole by the analogue beamformers, when tracking source 3C444. The figure shows the difference between the required delay (determined by the dipole's location and station pointing direction) and the actual delay that can be provided by the nearest beamformer setting. There is no pattern to the sign or magnitude of the errors.}
\label{fig:delayerrors}
\end{center}
\end{figure}

%- table of MWA dipole expected performance (A/T or SEFD) between 60 and 300 MHz.\\
%- loss of sensitivity due to imperfect delay settings
The maximum delay error on any dipole is half of the 435\,ps delay quantum of the first stage beamformers (the Kaelus second stage beamformer has a much smaller 92\,ps delay quantum), hence is 218\,ps or 6.5\,cm.
The actual errors will be uniformly randomly distributed between -218\,ps and 218\,ps depending on the pointing direction and location of the dipole in the array. An example is shown in Fig. \ref{fig:delayerrors}.
This translates to a standard deviation in phase error over all dipoles of 13, 6.5 and 4.4 degrees at 300, 200 and 100\, MHz respectively.
% stddev of uniformly distributed numbers between -1 and 1 is 0.56.
This phase error is comparable to the residual phase error from cable length differences, hence we combine the errors in quadrature to estimate the overall standard deviation in phase error to be 8.2 degrees at 200\,MHz. This phase error will cause at most 2\% sensitivity loss at 200\,MHz with smaller losses at lower frequencies.

% See eda_sensitivity_for_paper_COSMETICS.odt page 19 for input to this table:
% cd
% ~/Desktop/EDA/data/sensitivity/20161207_eda_to_mwa/3c444_evening/full_band_sensitivity
% interpolate_cubic.py eda_sensitivity_3C444_20161208_NoBeamformingErrors_TrcvFromSkyModelWithErrLimit50K_XX.txt  60 230 --y_axis_label="Sensitivity [m^2/K]" --outfile=eda_sensitivity_3C444_20161208_NoBeamformingErrors_TrcvFromSkyModelWithErrLimit50K_XX_SENSITIVITY_CUBIC.txt --coly=1  --step=10
% Single Dipole data from /home/msok/Desktop/EDA/loogbook/paper/single_mwa_dipole_aeff.odt  page 2
\begin{table*}
\begin{center}  
\begin{tabular}{@{}ccccc@{}}
\hline%
 Frequency [MHz]  & Sensitivity [$\mathrm{m}^2/\mathrm{K}$] & Effective Area [$\mathrm{m}^2$] &  \begin{tabular}{@{}c@{}}  System Temperature [K] \\ $\pm 10\%$ \end{tabular} & \begin{tabular}{@{}c@{}} Effective Area of a \\ single MWA dipole [$\mathrm{m}^2$] \end{tabular} \\
\hline
60 & 0.21 & 970 & 4968 & 14.68 \\
70 & 0.32 & 950 & 3218 & 10.72 \\
80 & 0.43 & 914 & 2156 & 8.16 \\
90 & 0.57 & 874 & 1538 & 6.39 \\
100 & 0.72 & 832 & 1150 & 5.13 \\
110 & 0.87 & 771 & 885 & 4.20 \\
120 & 1.01 & 707 & 698 & 3.49 \\
130 & 1.13 & 638 & 564 & 2.94 \\
140 & 1.22 & 568 & 465 & 2.50 \\
150 & 1.28 & 498 & 387 & 2.15 \\
160 & 1.34 & 435 & 325 & 1.86 \\
170 & 1.38 & 377 & 274 & 1.62 \\
180 & 1.35 & 329 & 244 & 1.42 \\
190 & 1.28 & 288 & 224 & 1.25 \\
200 & 1.21 & 252 & 206 & 1.10 \\
210 & 1.16 & 222 & 189 & 0.97 \\
220 & 1.14 & 196 & 173 & 0.87 \\
\hline
\end{tabular}
\end{center}
\caption{Expected performance of the EDA (the sensitivity values are the same as the blue curve in Figure~\ref{fig:eda_sensitivity} with the EDA pointed at 3C444; the sensitivity of the zenith-pointed array is approximately 5\% higher) and system temperature (black curve in Figure~\ref{fig:eda_t_sys}). The effective area of a single (standalone) MWA dipole was calculated in the same way as for the full EDA array, but with just a single dipole in the centre of the array (the results agree with a single MWA dipole over an infinite ground plane). The figures for system temperature are derived from the drift scans where sky model uncertainty dominates the error budget. We assign a conservative 10\% error to these. }
\label{tab_expected_performance}
\end{table*}

\subsection{Measured Performance}

\subsubsection{Drift scan observations}
\label{sec:drift_scan}

\begin{figure*}
\begin{center}
\includegraphics[width=1.0\textwidth]{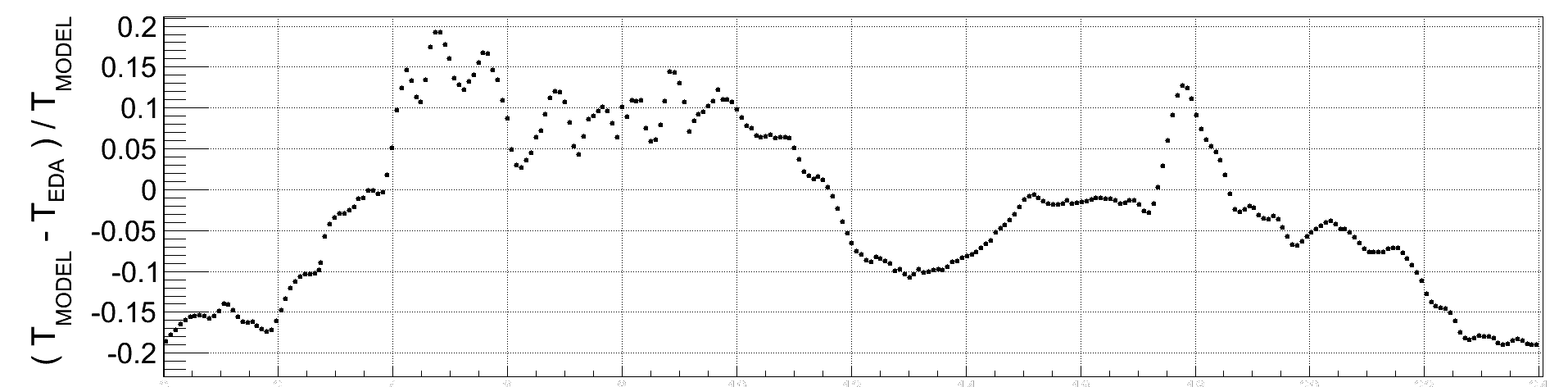}
\includegraphics[width=1.0\textwidth]{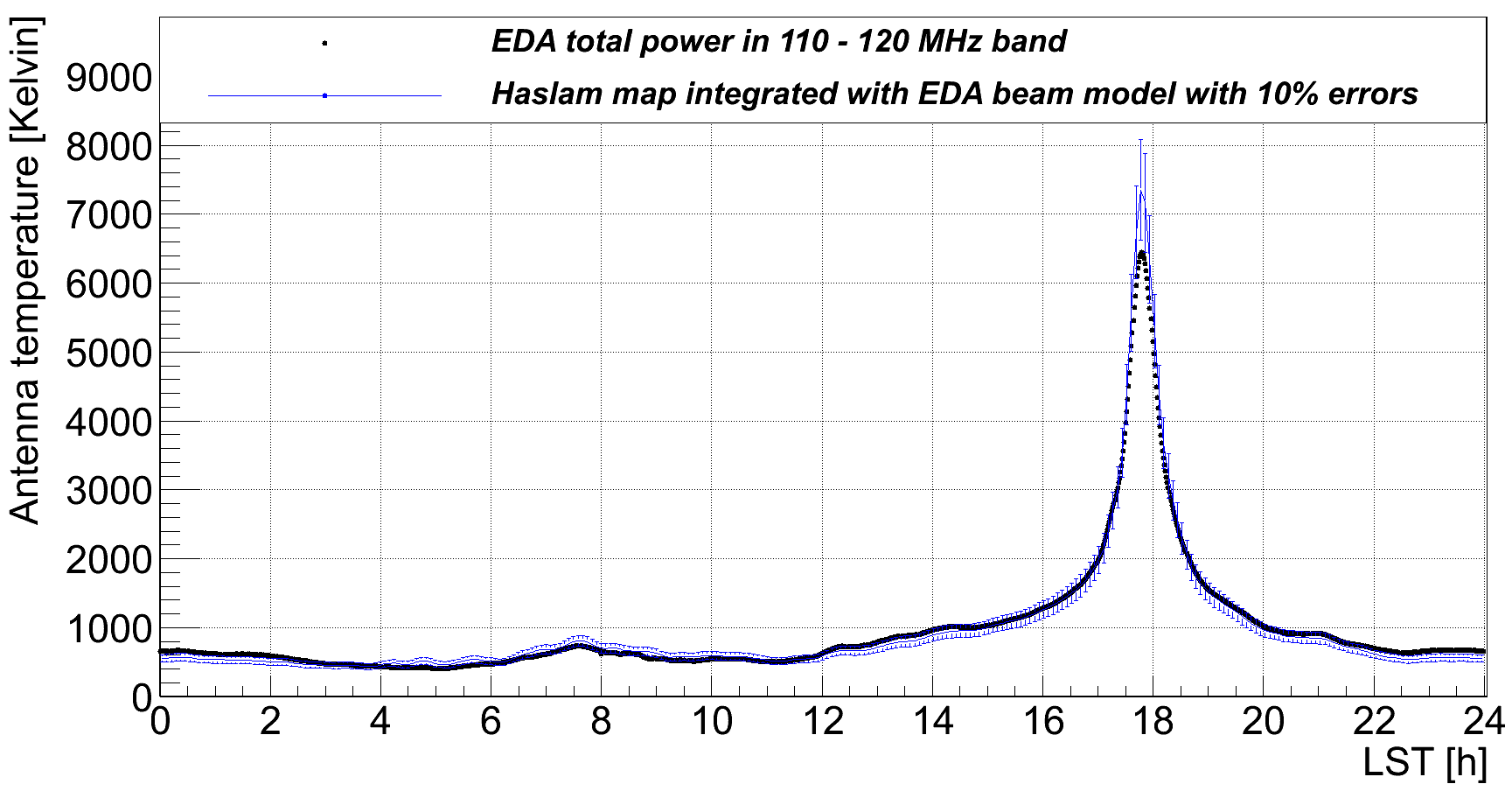}
\caption{Total power in the 110--120\,MHz band as a function of local sidereal time (LST) observed with the EDA pointed at zenith \newtext{(lower image)}. The black data points are the EDA data \newtext{(gain and receiver temperature were fitted in the entire 0-24\,h LST range)} and the blue curve with 10\% error bars are the EDA beam model data integrated with the measurements by \citet{1982A&AS...47....1H} at 408\,MHz scaled down to lower frequencies using a spectral index of $-2.55$. \newtext{The main source of uncertainty is the sky model. Therefore, we adopted a 10\% error as estimated by \citet{2008MNRAS.388..247D}, which was confirmed by the difference between the two curves with standard deviation of $\approx$10\% (upper image).} }
\label{fig:drift_scan}
\end{center}
\end{figure*}

In order to verify our EDA beam model we have collected several days of data in drift scan mode with the array pointing at zenith. 
The comparison of the total power collected in the 110--120\,MHz range with the model beam-weighted sky brightness temperature is shown in Figure~\ref{fig:drift_scan}.
The model is based on \citet{1982A&AS...47....1H} at 408\,MHz (the ``Haslam map'') scaled down to lower frequencies using a spectral index of $-2.55$, integrated with the EDA beam model at zenith.
The good agreement between model and data shows that the EDA beam forming was working as expected.
%(the gain and receiver temperature were fitted in the entire 0-24\,h local sidereal time (LST) range to scale the Haslam map data to the observed data). 

Encouraged by the good agreement between the data and the model, we used the drift scan observations to infer the receiver temperature of the EDA as a function of frequency.
\newtext{A similar approach of calibrating receiver temperature from beam and sky models was described in \citet{2004RaSc...39.2023R} and was  subsequently applied to MWA \citep{2007AJ....133.1505B}, PAPER \citep{2015ApJ...801...51J} and HERA\footnote{Memo 16 at http://reionization.org/science/memos/} instruments.}
We model the power $P(\nu)$ detected by the EDA at frequency $\nu$ as
\begin{equation}
P(\nu) = g(\nu)\biggl(T_{ant}^{model}(\nu) + T_{rcv}(\nu) \biggl),
\label{eq:power_vs_tmodel}
\end{equation}
where $g(\nu)$ is the gain of the EDA signal chain, $T_{rcv}(\nu)$ is the EDA receiver temperature and $T_{ant}^{model}(\nu)$ is the EDA antenna temperature.
The antenna temperature is calculated as the beam-weighted average sky temperature
\begin{equation}
T_{ant}^{model}(\nu) = \frac{\int_{4\pi} B(\nu,\theta,\phi) T(\nu,\theta,\phi) d\Omega}{\int_{4\pi} B(\nu,\theta,\phi) d\Omega}, 
\label{eq:sky_integration}
\end{equation}
where $B(\nu,\theta,\phi)$ is the EDA beam pattern (see Fig \ref{fig:EDA_beams}), $T(\nu,\theta,\phi)$ is the sky brightness temperature from the Haslam map at frequency $\nu$ and pointing direction $(\theta,\phi)$.

We observed that the measured power $P(\nu)$ versus predicted antenna temperature was extremely linear in LST range 13--17\,h where we avoid the Sun and Galactic Centre transits (the reliability of the Haslam map in the Galactic centre is poor due to the presence of \textsc{Hii} regions).
Therefore, we used least square fitting to obtain parameters $g(\nu)$ and $T_{rcv}(\nu)$ in every 10\,MHz frequency bin using the data in LST range 13--17\,h.
The receiver temperature $T_{rcv}(\nu)$ obtained with this method is shown in Figure~\ref{fig:eda_t_sys}.
We also tried using the Global Sky Model (GSM) of \citet{2008MNRAS.388..247D} and the results were very similar to those obtained using the Haslam map. 
\newtext{The results on $T_{rcv}(\nu)$ are in a very good agreement with figures (50 and 25\,K at 150 and 200\,MHz respectively) presented by \citet{2013PASA...30....7T} and recent laboratory measurements (Sutinjo et al. in preparation).}

% /home/msok/Desktop/EDA/loogbook/paper/T_rcv/eda_paper_lightcurve_and_trcv_FINAL.odt
% /home/msok/Desktop/EDA/data/2016-07/20160703/HASLAM/WithSun_and_BeamformingErrors/BIGHORNS/PAPER/images/final/total_power_110_120_MHz_vs_LST_DATA_and_MODEL.png
% cd /home/msok/Desktop/EDA/data/2016-07/20160703/HASLAM/WithSun_and_BeamformingErrors/BIGHORNS/
% root [1] .x fit_trcv_from_data_vs_time_NEW_binned_offset.C+("total_power_110_120_MHz_vs_LST.txt","total_power_110_120_MHz_MODEL/tant_vs_lst_1467504705_300sec_115MHz.out","power",115,"images/optimal_range/",0,24,1,"trcv_vs_freq_optimal_range_RAW.txt",0)
% 
% residuals :
% /home/msok/Desktop/EDA/data/2016-07/20160703/HASLAM/WithSun_and_BeamformingErrors/BIGHORNS/PAPER/images/final/total_power_110_120_MHz_vs_LST_DATA_and_MODEL_RESIDUALS_NORMALISED.png
% cd /home/msok/Desktop/EDA/data/2016-07/20160703/HASLAM/WithSun_and_BeamformingErrors/BIGHORNS
%  .x plot_trcv_from_data_vs_time_NEW_binned_offset.C+("total_power_110_120_MHz_vs_LST.txt","total_power_110_120_MHz_MODEL/tant_vs_lst_1467504705_300sec_115MHz.out","power",115,"images/optimal_range/",0,24,1,"trcv_vs_freq_optimal_range_RAW.txt",0)

\subsubsection{Measured sensitivity}
\label{sec:measured_sensitivity}

The sensitivity of the array
%,$S = A/T_{sys}$, where A is the effective area and $T_{sys}$ is the system temperature,
was measured by observing a strong compact source (3C444) close to its transit and calculated using the method described in \citep{2015ITAP...63.5433S}. 
Our model of the calibrator source 3C444 is derived from images from the Very Large Array Low-frequency Sky Survey Redux (VLSSr) at 74\,MHz \citep{2014MNRAS.440..327L}, where the higher frequencies are modelled by a power-law scaling calculated from the total flux density measured at 74\,MHz and 1.4\,GHz \citep{1998AJ....115.1693C}.
% \emph{We should include the model of 3C444's flux density vs frequency. Is it $S(\nu) = 79.6 (\nu/160\mathrm{MHz})^{-0.88}$\,Jy?}
The observations were performed when the MWA was in the compact configuration (with a maximum baseline of 1\,km and typical baseline length of 300\,m) and therefore the calibrator source remained unresolved in these observations on all baselines.

The system equivalent flux density (SEFD) of an antenna or station is directly measurable from calibrated visibilities.
With the EDA replacing one of the MWA's tiles, we recorded visibility data formed by the hybrid MWA-EDA array.
We used 24\,sec observations of 3C444 collected on 2016-12-07 between 10:24:40 and 10:32:00 UTC when 3C444 was at $(az,za)$ $\approx$ ($300\degree,18\degree$) where $az$ is the azimuthal angle measured east of north, and $za$ the zenith angle.
The sensitivity was calculated as $A/T_{sys} = 2k/SEFD$, where $k$ is Boltzmann's constant.

\begin{figure}
\begin{center}
\includegraphics[width=\columnwidth]{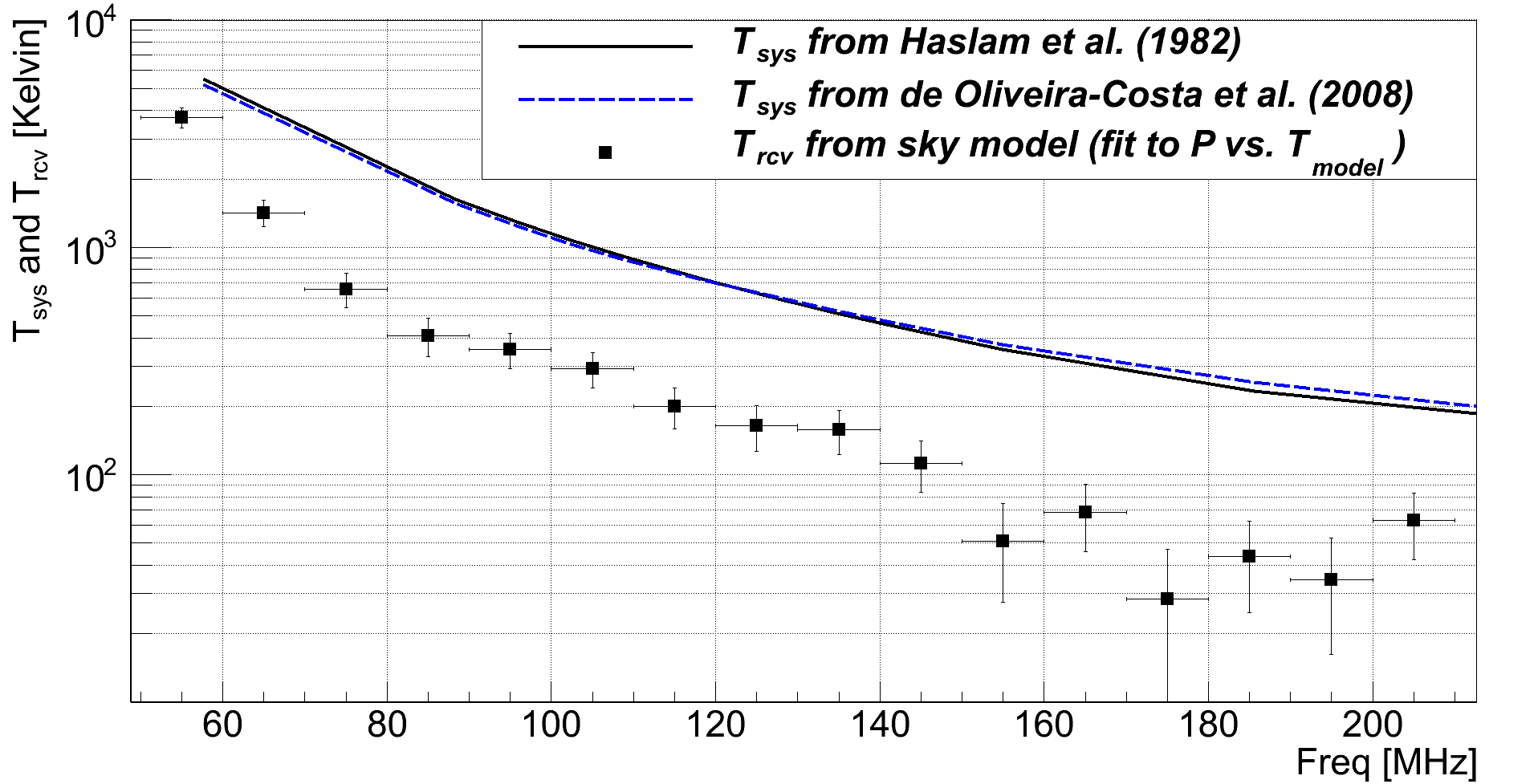}
\caption{Model system temperature, $T_{sys} = T_{ant} + T_{rcv}$, used in calculation of expected sensitvity (Fig.~\ref{fig:eda_sensitivity}). The black solid line was derived from $T_{ant}$ calculated using measurements by \citet{1982A&AS...47....1H} at 408\,MHz scaled down to low frequencies using a spectral index of $-2.55$ and integrated with the EDA beam (eq.~\ref{eq:sky_integration}) in the direction of 3C444. The blue dashed line was calculated using the Global Sky Model (GSM) of \citet{2008MNRAS.388..247D}. The black data points are $T_{rcv}$ derived from the sky model (see Sec.~\ref{sec:drift_scan} for details). Since system temperature is dominated by $T_{ant}$ derived from the sky models, 10\% error can be assumed \newtext{as typically quoted uncertainty of the sky models \citep{2008MNRAS.388..247D}}.
%\textbf{NEEDS UPDATE OR REMOVAL : The red dotted line is the measured noise temperature of the EDA receiver $T_{rcv}$ and the green line is $T_{rcv}/M$.}
}
\label{fig:eda_t_sys}
\end{center}
\end{figure}

\begin{figure}
\begin{center}
\includegraphics[width=\columnwidth]{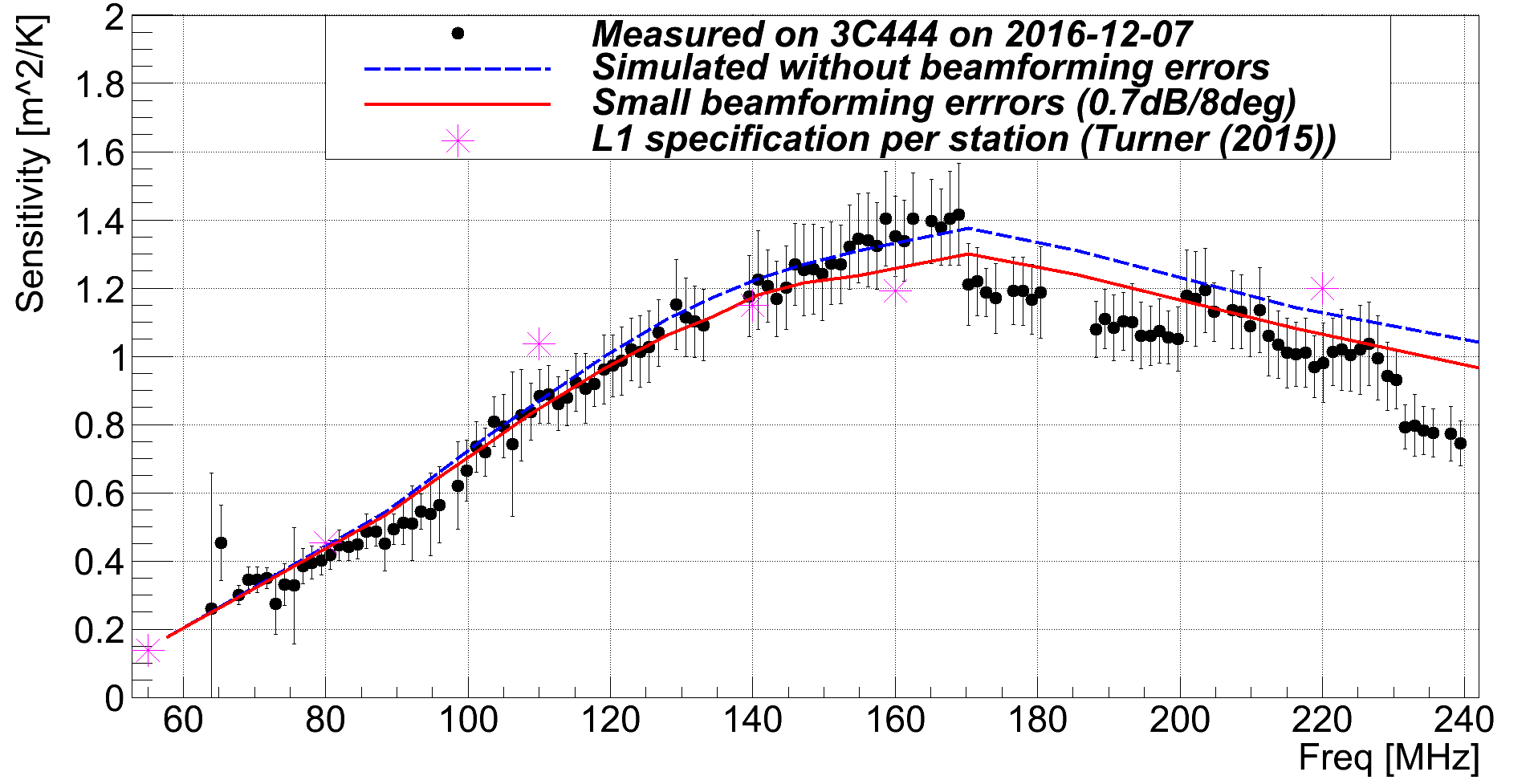}
\caption{The sensitivity ($A/T_{sys}$) of the EDA measured from 24\,sec observations of 3C444 at (az,za) $\approx($300\degree,18\degree$)$ between 10:24:40 and 10:32:00 UTC on 2016-12-07 (black data points). The blue dashed line is an expected ``ideal'' sensitivity of the array (without taking into account any degradation in beamforming performance). The red line is an expected sensitivity of the array with degradation due to second stage beamforming taken into account (standard deviation of random variations in amplitude and phase of individual dipoles being $0.7$\,dB and $8\degree$ respectively).
\newtext{The magenta asterisks are the SKA Phase 1 specifications as defined in revision 10 of \citet{Turner2015}.}
The abrupt drop in measured sensitivity between 170 and 200\,MHz, and above 230\,MHz is an artefact of in-band radio frequency interference (RFI).
}
\label{fig:eda_sensitivity}
\end{center}
\end{figure}

The measured sensitivity in the E-W polarisation is shown in Figure~\ref{fig:eda_sensitivity} together with the expected sensitivity at the same pointing direction and time.
The expected sensitivity, $A/T_{sys}$, was derived using the EDA station beam model to calculate the effective area $A$ in the direction of the calibrator source at the time of the observations.
The effective area calculated every 10\,MHz is given in Table~\ref{tab_expected_performance} and it can be calculated by multiplying the sensitivity in Figure~\ref{fig:eda_sensitivity} by the system temperature at the corresponding frequency (black curve in Fig.~\ref{fig:eda_t_sys}). Note that the calculated effective area is for the off-zenith pointing at 3C444, hence will be approximately 5\% less than the zenith-pointed effective area.

The system temperature was calculated as $T_{sys} = T_{ant} + T_{rcv}$, where $T_{rcv}$ is the noise temperature of the EDA receiver chain (LNAs and beamformers) and $T_{ant}$ is the antenna temperature.
The receiver noise temperature $T_{rcv}$ was measured from the drift scan data with the EDA beam pointed at zenith (Sec.~\ref{sec:drift_scan}).
%\textbf{ADD: possibly comparison with Adrian/Budi.}
% Adrian/Budi version 
% The system temperature was calculated as $T_{sys} = T_{ant} + T_{rcv}/M$, where $T_{rcv}$ is the noise temperature of the EDA receiver chain (LNAs and beamformers), $T_{ant}$ is the antenna temperature, $M=(1-|\Gamma_a|^2)$ and $\Gamma_a$ is the reflection coefficient of the antenna.
% The receiver noise temperature $T_{rcv}$ measured for the EDA receiver ranges from $\sim$130\,K at 60\,MHz to about 20--60\,K between 100--300\,MHz and $T_{rcv}/M$ is considerably higher especially at lower frequencies (Fig.~\ref{fig:eda_t_sys}).

The antenna temperature, $T_{ant}$, was calculated as the beam-weighted sky brightness temperature of the Haslam map, as described above.
We have also tested calculation of $T_{ant}$ using the Global Sky Model (GSM) of \citet{2008MNRAS.388..247D}, which led to a system temperature higher by 10--20\,K above 150\,MHz (Fig.~\ref{fig:eda_t_sys}), resulting in a small ($\sim 0.1$\,$\mathrm{m}^2/\mathrm{K}$) reduction in inferred sensitivity at these frequencies.
In these calculations we used an analytical beam model of the EDA, but we have also tested an implementation of the full FEKO\footnote{\url{https://www.feko.info/}}-based beam model \citep{mwa_fullee_msok} for the EDA and the difference in expected sensitivity was insignificant.

The theoretically expected sensitivity of the EDA is shown in Figure~\ref{fig:eda_sensitivity} as blue and red curves. 
The blue curve was calculated for an ``ideal'' case (without any degradation in beamforming performance) and the red curve was calculated taking into account the degradation in beamforming performance due to random variations in amplitude and phase of individual beamformers with standard deviations of $0.7$\,dB and $8\degree$ respectively (Sec. \ref{sec:expectedperf}).
The gain variations represent realistic values that were measured for EDA components prior to deployment and confirmed by measurements of sky power variations using all 256 individual dipoles.
The variations of phases between the 16 beamformers were measured using a strong calibrator source (Sec.~\ref{sec:performance}).

The abrupt drop in measured sensitivity between 170 and 200\,MHz is an artefact due to in-band radio frequency interference (RFI) that was present during the observation of that frequency range\footnote{The MWA typically observes in contiguous blocks of frequency that are 30.72\,MHz wide. The 170--200\,MHz block is one of three that are commonly used to cover the 140--231\,MHz range.}.
Although we have excised affected data using the standard MWA pre-processing tools \citep{2015PASA...32....8O}, it is clear that remaining low-level RFI is affecting the sensitivity measurement.
\newtext{The low-level RFI in 170--200\,MHz band was due to digital TV signals (also observed by BIGHORNS instrument \citep{2015ApJ...813...18S}) from remote transmitters propagated via reflection or tropospheric ducting effects. The tropospheric ducting phenomena at the MRO is more common in the summer time \citep{bighorns_ducting} when the sensitivity measurements were performed. Similarly, the small deviation from expected sensitivity in $\approx$88--100\,MHz band may also be due to residual RFI in the FM band (88--108\,MHz).}
The range between 230 and 240\,MHz is similarly affected by persistent satellite-based RFI that is always present above approximately 240\,MHz.
Absent of RFI, we expect the measured sensitivity to follow the red curve on Figure \ref{fig:eda_sensitivity} as it does in the remainder of the frequency range.

% NEWEST (2017-04-06) - red and green curves removed to only keep sky model based measurements. Ill. 14 in /home/msok/Desktop/EDA/loogbook/paper/T_rcv/eda_paper_lightcurve_and_trcv_FINAL.odt
%   Ill. 14 in /home/msok/Desktop/EDA/loogbook/paper/T_rcv/eda_paper_lightcurve_and_trcv_FINAL.odt
%   /home/msok/Desktop/EDA/data/sensitivity/20161207_eda_to_mwa/3c444_evening/full_band_sensitivity/images/final/paper/FINAL/20170406/Tsys_HASLAM_ANGELICA_TrcvDivM_FromSky.png

%%%%%%%%%%%%%%%%%%%%%%%
%%%%%%%%%%%%%%%%%%%%%%%
\section{DISCUSSION}
\label{sec:discussion}
%- uncertainty in sky model directly translates to uncertainty in sensitivity\\
%- derivation of Trec using drift scans\\
The uncertainty in the derived sensitivity of the EDA is dominated by the uncertainty in $T_{rcv}$, which is in turn dominated by uncertainty in the global sky model extrapolated down to low frequencies.
However, the very good match between the predicted sensitivity (which used the derived receiver temperature from drift scans) and the measured sensitivity from the calibrator source (Fig. \ref{fig:eda_sensitivity}) provides confidence that the method used to derive $T_{rcv}$ is appropriate.

Fig. \ref{fig:eda_t_sys} shows the EDA $T_{rcv}$ is typically 20\% of $T_{sky}$ over most of the measured frequency range and is closer to 10\% of $T_{sky}$ in the key frequency range for Epoch of Reionisation (EoR) science (150--200\,MHz).
A key requirement of low frequency arrays is to be sky-noise limited, and the EDA is clearly so over most of the measured range.
At the lowest frequencies (less than 60\,MHz) the receiver temperature is approaching the sky temperature, so the array is not sky-noise limited at high Galactic latitudes, but it may still be sky-noise limited on the Galactic plane.
The frequency range between approximately 240 and 290\,MHz is inaccessible due to persistent satellite interference.
We did not measure the performance of the array at higher frequencies (above 300\,MHz) but we would expect the performance to be dominated by the LNA- antenna matching and dipole effective area.

%Response of MWA dipole is relatively poor above 300\,MHz. Usable but not sky noise limited below 60\,MHz.\\
%- overall excellent agreement between predicted sensitivity and measured\\

\subsection{Revisiting the goals of the EDA}
The EDA is an SKA-Low sized station whose primary antenna elements are MWA dipoles and beamformers.
It was conceived of, designed, and deployed in 18 months.
This rapid deployment timescale was made possible by re-use of well understood MWA components and technology.
The modular nature of the EDA means that it is possible and straightforward to upgrade parts of the system without affecting the rest of the array.

Over the frequency range where measurements are possible, the EDA performs as expected from basic antenna theory with the overall receiver temperature determined by the properties of the MWA dipole (including LNA).
The EDA is thus a well understood reference against which comparisons of prototype SKA-Low hardware, or alternate SKA-Low designs, can be made.

The EDA demonstrates that first stage analogue beamforming is a viable method to form full station beams with full sensitivity. Hence, analogue beamforming presents an option to reduce cost, risk, and complexity in the case where SKA-Low is faced with budgetary or technology readiness pressures.

The MWA's correlator was enhanced to be able to accept digital data streams from external instruments in anticipation of future cross-correlation of the MWA with EDA and AAVS1 signals. This process uncovered an unexpected issue associated with cross correlating digital datastreams formed by heterogeneous digital filterbanks, the details of which are beyond the scope of this paper. Although we were able to proceed with the sensitivity tests by connecting the analogue signal from the EDA directly to an MWA receiver, the many benefits of early prototyping, both for direct experience building a system and to uncover unexpected problems, have been borne out again by the EDA.

\subsection{Station and dipole element effective area}
%- significant difference in effective area of dipole standalone vs in array. Overall performance is determined entirely by array layout.\\
The layout of the station was designed (based on SKA specifications) to make the transition from dense array to sparse array around 110\,MHz.
This transition can be seen in the centre and right panels of Figure \ref{fig:EDA_beams} with the appearance of a ring of relatively high power, large scale sidelobes for the sparse arrays at 200 and 300\,MHz.

We have included in Table \ref{tab_expected_performance} the theoretical effective area of a standalone MWA dipole over groundscreen to highlight the  reduction in actual effective area that a dipole suffers when placed in an array. While this is obviously true at lower frequencies where the array layout is dense (defined as where the average spacing between dipoles less than half the wavelength), it can also be true in the regime where the array would be considered sparse.
For example, at 150\,MHz a standalone MWA dipole has effective area of 2.1\,m$^2$ whereas the average effective area per dipole in the array is 1.9\,m$^2$.
This effect is almost entirely determined by the layout of the station and hence calculation of theoretical array sensitivity based on raw dipole sensitivity alone, without considering the effects of the array factor, are likely to be misleading.

%- discussion of dense-sparse transition\\
%- simulated beam patterns at 100, 200 and 300 MHz.\\

%%%%%%%%%%%%%%%%%%%%%%
%%%%%%%%%%%%%%%%%%%%%%
%\section{THE FUTURE: UPGRADED BEAMFORMERS AND DIGITAL SECOND STAGE BEAMFORMING}
\section{THE FUTURE}
%- plans for future\\

\subsection{Digital second-stage beamforming}
The longer term plan for the EDA is to replace the analogue second stage beamformer with systems that digitise the data after first stage beamforming and combine the data digitally.
Such an architecture is similar to the LOFAR High Band Antenna (HBA) station \citep{2013A&A...556A...2V}, although the EDA does not have the standard size tiles of the LOFAR-HBA or the MWA.
The proposed architecture follows a well established design consisting of:
\begin{itemize}
\item digitising the entire signal between 0 and approximately 350\,MHz;
\item applying a delay correction for each sub-array by aligning the datastreams to the nearest integer number of samples;
\item channelising the signal into $\sim 1$\,MHz ``coarse'' channels using an oversampled polyphase filterbank (PFB);
\item applying a calibration gain and sub-sample delay correction (phase shift) in the coarse channels;
\item selecting a subset of these channels to form a station beam (the total bandwidth can be scaled depending on network/compute resource limitations);
\item forming a station beam by adding the data from all sub-arrays;
\item transmitting the station beam to downstream devices (e.g. a pulsar backend or a correlator) via a standard network interface.
\end{itemize}

Implied in this architecture is another ``fine'' channelisation in downstream processing systems to the frequency resolution required by the science program.
The fine channelisation is required in a downstream correlator so that overlapping frequency ranges from the coarse PFB can be discarded and a continuous smooth broad frequency range can be recovered by stitching fine channels together.

We also note that an alternate architecture is conceivable where the first filterbank transforms the data directly to fine channels using a filterbank with a large number of channels. This approach has some advantages, however the FPGAs used in present implementations of these digital systems are not well suited to doing this, and transforms are typically limited to approximately 4096 channels.

A number of off-the-shelf solutions for digitising the data and performing the coarse channelisation exist.
As with the initially deployed system, the goal is to use existing hardware and software where possible and to avoid requiring specialist expertise in FPGA programming.
The CASPER\footnote{\url{http://casper.berkeley.edu/}} toolset provides a potential solution and has existing libraries to implement digital filterbanks and data transport.
We are also investigating the USRP and FlexRio platforms from Ettus Research and National Instruments respectively.
These systems have the benefit of being able to be programmed using the high-level LabVIEW language and have optimised pre-existing libraries for complex and/or compute-intensive tasks.

%We have already shown, both in \citet{2015ITAP...63.5433S} and this work, that the sub-arrays formed by the 16 dipoles with a first stage beamformer have sufficient sensitivity and a sufficiently small beam area, that high precision gain and delay measurements are straightforward to obtain using bright compact astronomical sources.

\subsection{Upgraded beamformers}
The identification of the detrimental effects to EoR science of rapidly varying (with frequency) bandpass ripple in the MWA \citep[e.g.][]{2016MNRAS.460.4320E,2016MNRAS.458.1057O,2016ApJ...833..102B} has become a significant new design constraint \citep{2016PASA...33...19T,2016MNRAS.461.3135B,2016ApJ...826..199N} for SKA-Low, HERA and future upgrades of MWA hardware.
The EDA will be used to prototype and verify upgraded MWA beamformers that use optical RF-over-fibre outputs, rather than coaxial cable, which show promising spectral smoothness in laboratory tests when coupled with appropriate optical isolators.
The upgraded beamformers (with optical output signal) could also be connected to the current prototype SKA-Low Tile Processing Module (TPM) optical inputs.

%%%%%%%%%%%%%%%%%%%%%%%
%%%%%%%%%%%%%%%%%%%%%%%
\section{CONCLUSION}
\label{sec:conclusion}
We have described the design and measured performance of the Engineering Development Array (EDA).
It is a station of 256 antennas, spread over a diameter of 35\,m, formed by 16 groups of 16 pseudo-randomly placed MWA dipoles (Fig. \ref{fig:dipole_layout}).
The EDA re-uses as much MWA hardware and software as possible. It uses standard MWA analogue beamformers to perform first stage beamforming (hence the EDA is comprised of 16 sub-arrays) and standard MWA dipoles with slightly modified LNAs to receive signals down to 50\,MHz.

The EDA was conceived as a rapidly deployable test and verification system to support the development, test and verification programs for both SKA Low and MWA. In its initial incarnation, the EDA uses two-stage analogue beamforming to form an array beam, however work is underway to upgrade to digital second stage beamforming.

Using drift scans and a model for the sky brightness temperature at low frequencies, we have derived the EDA's receiver temperature as a function of frequency. The results show that the EDA is sky-noise limited over most of the frequency range measured, with the exception being below 60\,MHz where the receiver temperature is comparable to the sky temperature at high Galactic latitude.

Using the derived receiver temperature we have measured the sensitivity of the array by measuring the noise variance in calibrated visibilities.
The measured sensitivity agrees very well with the predicted sensitivity of the array taking into account losses due to uncorrected gain errors in the second stage beamformer. The results demonstrate the practicality and feasibility of using MWA-style precursor technology for SKA-scale stations and highlight the benefits of rapid prototyping and verification for array development.

\begin{acknowledgements}
This scientific work makes use of the Murchison Radio-astronomy Observatory, operated by CSIRO. We acknowledge the Wajarri Yamatji people as the traditional owners of the Observatory site. Support for the operation of the MWA is provided by the Australian Government (NCRIS), under a contract to Curtin University administered by Astronomy Australia Limited. We acknowledge the Pawsey Supercomputing Centre which is supported by the Western Australian and Australian Governments.

This research was supported under the Australian Research Council's Discovery Early Career Researcher funding scheme (project number DE140100316), and the  Centre for All-sky Astrophysics (an Australian Research Council Centre of Excellence funded by grant CE110001020).
\end{acknowledgements}

\nocite*{}
\bibliographystyle{pasa-mnras}
\bibliography{refs}

\end{document}